\documentclass[pra,twocolumn,showpacs,preprintnumbers,amsmath,amssymb,superscriptaddress]{revtex4}

\usepackage{graphicx}
\usepackage{dcolumn}
\usepackage{bm}

\begin{document}

\title{
Relativistic many-body calculation of low-energy dielectronic resonances in
Be-like carbon}

\author{A. Derevianko}
\affiliation {Department of Physics, University of Nevada, Reno, Nevada 89557}

\author{V. A. Dzuba}

\affiliation{School of Physics, University of New South Wales,
        Sydney, 2052, Australia}
\affiliation{Department of Physics, University of Nevada, Reno, Nevada 89557}

\author{M. G. Kozlov}
\affiliation{Petersburg Nuclear Physics Institute, Gatchina 188300, Russia}
\affiliation{Department of Physics, University of Nevada, Reno, Nevada 89557}

\date{\today}

\begin{abstract}
We apply relativistic configuration-interaction method coupled with many-body
perturbation theory (CI+MBPT) to
describe
low-energy dielectronic
recombination. We combine the CI+MBPT approach with the complex rotation
method (CRM) and compute the dielectronic recombination spectrum for Li-like
carbon recombining into Be-like carbon. We demonstrate the utility and
evaluate the accuracy of this newly-developed CI+MBPT+CRM approach by
comparing our results with the results of the previous high-precision study of
the CIII system [Mannervik et al., Phys.\ Rev.\ Lett.\ {\bf 81}, 313 (1998)].
\end{abstract}

\maketitle

\section{Introduction}

One of the important atomic processes governing ionic charge
abundances in plasmas is dielectronic recombination (DR). The DR
process is a two-stage reaction with a formation of an intermediate doubly
excited ion and a subsequent relaxation via photon emission,
\begin{equation}
{e^ - } + {A^{q + }} \to {\left[ {{A^{(q - 1) + }}} \right]^{**}}
\to {\left[ {{A^{(q - 1) + }}} \right]^*} + {\rm{photon}} \, .
\label{Eq:Reaction}
\end{equation}
Due to the importance of DR in plasma processes, there has been a large
body of
systematic
experimental and theoretical work on DR. The present status of the field is
reviewed in Ref.~\cite{KalPal07}.

An excellent review of current theoretical methods of treating DR may be found
in Refs.~\cite{PinGriBad05,KalPal07}. DR calculations were carried
out using configuration-interaction (CI),
multi-configuration Hartree-Fock (MCHF), and techniques of many-body-perturbation theory (MBPT) with
the help of standard codes, such as GRASP \cite{ParFroGra96}, CIV3
\cite{Hib75}, MCHF code \cite{FisBraJon97}, AUTOSTRUCTURE~\cite{Bad86}, and
others. For electron temperatures
$T_e \gtrsim 100$ eV,  there is  a
good agreement between calculated and measured DR rates. However,
for $T_e \lesssim 10$ eV there are significant disagreements between theory
and experiment  (see, e.g, Refs.~\cite{SavKahGwi03,SchSchBra04,KalPal07}).
These discrepancies are usually attributed to theoretical inaccuracies in the
positions of low-energy
resonances $E_r < 1$ eV. Even small shifts of such resonances to the lower
energies lead to underestimating  the DR rate~\cite{SchSchBra04}.

The DR process is a resonant process: cross-section spikes at electron kinetic
energies
that are resonant with internal transitions between bound ionic
states. As a result, the  DR rate coefficients, entering, e.g., plasma
ionization stage calculations, are exponentially sensitive to uncertainties in
energies of resonances, $E_r$.
Because of this exponential sensitivity,
there is an outstanding and practically-relevant problem:
a reliable description of the DR at low temperatures. This problem has been
highlighted, for example, by Savin {\it et al.}~\cite{SavGwiGri06}. These
authors write, {\em ``the single greatest challenge facing modern DR theory is
accurately calculating the resonance structure for the low collision energies
needed to calculate low-temperature DR.''}~\footnote{This statement is somewhat
softened when the Rydberg states are populated in relatively dense plasmas (F. Robicheaux, S. D. Loch, M. S. Pindzola,
C. P. Ballance, private communications).} Compared to high energies (where a
simplified Rydberg-like description suffices), at low excitation energies the
positions of involved atomic resonances become sensitive to {\em many-body
correlations}.  Solving the correlation problem accurately is a challenging
task, and the existing approaches have difficulties in reliably describing
the low-temperature DR.

The most accurate method to date in treating the low-temperature DR is the
relativistic
many-body theory by the Stockholm group (see, e.g.,
\cite{Lin94,ManDeWEng98,LinDanGla01,TokEklGla02} and references therein).
Our present approach  shares essential elements
with this
highly-successful method: although our computational toolbox is different, it
is also
based on the many-body theoretical treatment and it is {\em ab initio}
relativistic. There is, however,
an important difference: all the enumerated calculations by  the Stockholm
group have final ions, $A^{(q-1)+}$ with two electrons outside a closed-shell
core. Our methodology is more general and allows one to  investigate systems with
a larger number of valence electrons. The goal of the present paper is to
evaluate the utility and the accuracy of
our approach by comparing our results with the benchmark data of
Ref.~\cite{ManDeWEng98}.

The paper is organized as follows. In Section~\ref{Sec:Method} we present a
discussion of
basic formulas of DR (Section~\ref{Sec:DRformulas}), the CI+MBPT approach and
the complex rotation method. Specifics of the calculations and numerical
results for Be-like carbon are presented
in Section~\ref{Sec:C3plus}.

\section{Method}
\label{Sec:Method}

\subsection{Dielectronic recombination}
\label{Sec:DRformulas}

We start by formalizing the DR reaction, Eq.~(\ref{Eq:Reaction}), and
recapitulating well-known results for the DR cross-section and rate
coefficient. In an independent particle picture, the incident electron of
energy $\varepsilon$ excites one of the bound electrons of the $A^{q+}$ ion
($n_a \ell_a \rightarrow n_b l_b$) and at the same time the initially free
electron is captured into one of the excited orbitals $n\ell$ of the target
ion: this forms a doubly-excited   ion $\left[ {{A^{(q - 1) + }}}
\right]^{**}$. This intermediate step is concluded by a radiative decay to a
final state below the ionization threshold. The theoretical description
requires two distinct ingredients: capture (auto-ionizing) amplitudes due to
electron-electron interactions and transition amplitudes due to the photon
bath. In the isolated-resonance approximation (valid for the commonly
encountered situation when there are no overlapping resonances of the same
symmetry), the DR cross-section due to an individual resonance may be
parameterized by the Lorentzian~(see, e.g., \cite{TokEklGla02})
\begin{equation}
 {\sigma _r}\left( \varepsilon  \right)
 = S_r \times \frac{1}{\pi }\frac{{{\Gamma_r}/2}}
 {{{{\left( {\Delta {E_r} - \varepsilon } \right)}^2}
 + {{\left( {{\Gamma_r}/2} \right)}^2}}} \, .
\end{equation}
Here $r$ labels the specific resonant state, $\Delta {E_r}$ is the energy of
the resonance with respect to the initial state $|i\rangle$ of the $A^{q+}$
ion, and $\Gamma_r$ is the width of the resonant metastable state. For a given
range of incident electron energies, the total cross-section is obtained by
summing over the resonances falling within that range. The strength, $S_r$, is
given by
\begin{equation}
S_{r}=\frac{\pi^2 \hbar ^{3}}{2m_{e} \,
\Delta E_{r}}\frac{g_{r}}{g_{i}} \,
\frac{A_{i\rightarrow r}^{a}~A_{r}^{rad}}{A_{r}^{a}+A_{r}^{rad}} \, .
\label{Eq:S}
\end{equation}
Here $g_r$ and $g_i$ are the multiplicities of the resonant and the initial
ionic states.  The radiative decay rate from the resonant doubly-excited
state, $A_{r}^{rad}$, is the conventional Einstein $A$ coefficient for
spontaneous emission summed over all possible final recombined states.
Usually, limiting the radiative decay channels to the electric-dipole
transitions is sufficient.
Finally, $A_{i\rightarrow r}^{a}$ is the capture (inverse of the Auger process)
rate to the resonant doubly-excited state and $A^{a}_r$ is the total
autoionization rate. The autoionizing and radiative rates account
for the total width of the resonance, $\Gamma_r=A^{rad}_r+A^{a}_r$.

The above formulation is an approximation: it omits radiative recombination
(RR), possible effects of interference between the RR and DR amplitudes,
interference between nearby resonances, modification of the Lorentzian profile
in the vicinity of the threshold, etc. While being approximate, this
treatment, however, is known to be of a sufficient quality for practical
calculations~\cite{PinGriBad05}.

\subsection{CI+MBPT method}
\label{Sec:CIMBPT}

Our calculations are performed using a method which
combines the configuration interaction (CI) technique with  many-body
perturbation theory (MBPT).
MBPT provides a systematic prescription for solving
the atomic many-body problem~\cite{LinMor86}. Basically, the residual (i.e., beyond the mean-field,
Dirac-Hartree-Fock (DHF), potential) Coulomb
interaction
between the electrons is treated as a perturbation and one applies
machinery similar to the textbook stationary perturbation theory. The MBPT,
when accompanied by a technique of summing up important chains of higher-order
diagrams to all orders,
produces excellent results for  atoms and ions which have
only one electron outside closed shells (see, e.g.~\cite{DerPorBel08,Dzu08}). Even for an atom as heavy as neutral Cs (55
electrons), the modern {\em ab initio} many-body relativistic techniques
demonstrate an accuracy of $\sim$ 0.1\% for removal energies, hyperfine structure
constants, and lifetimes~\cite{PorBelDer09}.

For atoms with two or more valence electrons the MBPT has significant
difficulties. The dense spectrum of  many-valence-electron atoms leads to
small energy denominators and the perturbative treatment of the valence-valence
correlations breaks down. An adequate technique in
this case is the CI method. Combining CI and MBPT allows one
to treat both valence-valence and core-valence correlations to  high
accuracy.

The CI+MBPT technique  has been described in detail in our previous
works~\cite{DFK96b,DzuJoh98,KPJ01,Dzu05a,Der01}. Below we briefly reiterate its main features.
For simplicity, let us consider a system with two valence electrons outside a closed
shell core (e.g., a Be-like C$^{3+}$ ion).
The calculations start from the $V^{N-2}$ approximation~\cite{Dzu05a}, which means
that the initial DHF procedure is carried out for the closed-shell ion, with the
two valence electrons removed.
The effective CI Hamiltonian for the divalent ion is the sum
of two single-electron Hamiltonians plus an operator representing
interaction between valence electrons:
\begin{equation}
  \hat H^{\rm eff} = \hat h_1(1) + \hat h_1(2) + \hat h_2(1,2).
\label{heff}
\end{equation}

The single-electron Hamiltonian for a valence electron has the form
\begin{equation}
  \hat h_1 = \hat h_0 + \hat \Sigma_1,
\label{h1}
\end{equation}
where $\hat h_0$ is the relativistic DHF Hamiltonian:
\begin{equation}
  \hat h_0 = c \mathbf{\alpha \cdot p} + (\beta -1)mc^2 - \frac{Ze^2}{r} + V^{N-2},
\label{h0}
\end{equation}
and $\hat \Sigma_1$ is the correlation potential operator describing
correlation interaction of the valence electron with the core \cite{DFK96b}.

Interaction between valence electrons is the sum of the Coulomb interaction
and two-particle correlation correction operator $\hat \Sigma_2$:
\begin{equation}
  \hat h_2 = \frac{e^2}{|\mathbf{r}_1 - \mathbf{r}_2|} + \hat \Sigma_2(1,2),
\label{h2}
\end{equation}
Qualitatively,
$\hat \Sigma_2$ represents the screening of the Coulomb interaction between
the valence electrons by the core electrons \cite{DFK96b}.

A two-electron wave function $\Psi$ for the valence electrons  has a form of
expansion over single-determinant wave functions
\begin{equation}
  \Psi = \sum_i c_i \Phi_i(1,2),
\label{psi}
\end{equation}
where $\Phi_i$ are the Slater determinants constructed from single-electron valence basis
states calculated in the $V^{N-2}$ potential.
The coefficients $c_i$ as well as two-electron energies are found by
solving the matrix eigenvalue problem
\begin{equation}
  (H^{\rm eff} - E)X = 0,
\label{Schr}
\end{equation}
where $H^{\rm eff}_{ij} = \langle \Phi_i | \hat H^{\rm eff} | \Phi_j \rangle$
and
$X = \{c_1,c_2, \dots , c_n \}$.

To calculate the correlation correction operators $\hat \Sigma_1$ and
$\hat \Sigma_2$ we use the second-order MBPT.

Technically, one needs a complete set of single-electron states to calculate $\hat \Sigma$
and to construct two-electron basis states for the CI
calculations.
To this end, we generate a finite basis set using B-splines and the dual-kinetic
balance method~\cite{BelDer08} for the $V^{N-2}$ DHF potential.

By diagonalizing the effective Hamiltonian we find the  wave functions;
they are further used for computing atomic properties such as electric-dipole transition
amplitudes. To compute matrix elements we
apply the technique of effective all-order (``dressed'') operators.
In particular, we employ the random-phase approximation (RPA). The RPA sequence
of diagrams describes a shielding of the externally applied field by the core
electrons.

The CI+MBPT calculations for a system with $n$ valence electrons
follow the very same scheme, with an expansion over Slater determinants
for $n$ electrons. Again the strong interaction between the valence electrons is
treated within CI, while the core-valence interactions are taken into account
in the MBPT framework (intermediate states in the operator $\Sigma$ include core excitations).

In principle, the effective Hamiltonian $H^{\rm eff}$ for systems with $n>2$
valence electrons includes three-particle operator $\hat{\Sigma}_3$, whose
computation is very costly. In Refs.~\cite{DFK96b,KPJ01} this operator was
calculated for neutral Thallium and respective contribution was found to be
negligible. One can expect that this conclusion should hold for all systems
with three-four valence electrons. For combinatorial reasons the relative role
of the three-particle interaction $\hat{\Sigma}_3$ rapidly grows with $n$ and
one may need to include it for systems with $n>4$.


\subsection{Complex rotation method}
\label{Sec:CRM}

Using the described CI+MBPT method we can find spectra of multi-valent ions.
This section connects the
CI+MBPT method to the DR problem. A straightforward computation within the
CI+MBPT has certain problems, discussed below. These problems, however, are
elegantly solved using the complex rotation method (CRM): a relatively minor
modification of the CI+MBPT method allows us to  compute positions
and widths of the dielectronic resonances.

What are the difficulties?\\
(i) The CI+MBPT method starts from a finite set of  single-particle states
(orbitals) computed in a spherical cavity of radius $R$.  The entire continuum
in practice is approximated by 20-30 orbitals per partial wave; their
individual energies depend on $R$. The DR doubly-excited resonance states are
embedded into the continuum. In some cases, depending on $R$, the resonance
state may become degenerate with the quasi-continuum states. This leads to a
requirement that the model CI space includes both the bound and the
quasi-continuum many-particle states (otherwise the perturbative treatment may break down due to small energy denominators). {\em How would
we separate the doubly-excited resonance states of interest from the
background quasi-continuum?}

(ii) Straightforward computation of the capture (Auger) rates starting from
the Fermi golden rule requires continuum wave-functions. Because we start with
the box quantization, we cannot easily generate the continuum orbitals of a
prescribed energy (in principle, this is possible by tuning the radius of the
cavity, but this is not a very practical solution). {\em How do we determine the
autoionizing rates without knowing the scattering states?}

Both difficulties are elegantly solved within the CRM
framework.

The complex-rotation method is well established and has been employed in
atomic physics and quantum chemistry for several decades (see, e.g., a
review~\cite{Rei82} and references therein). Previously, the CRM was
successfully applied to  the DR problem by E. Lindroth and collaborators
\cite{Lin94,ManDeWEng98,LinDanGla01,TokEklGla02}. In the CRM, the radial
coordinate is scaled by a complex factor $e^{i \theta}$
\begin{equation}
 r \to r e^{i \theta} \, , \label{Eq:rthetaScaling}
\end{equation}
$\theta$ being  an adjustable parameter. For example, the radial Dirac
equation becomes
\begin{align*}
&\left(  V\left( e^{i \theta}  r\right)  +c^{2}\right)  P_{n\kappa}\left(
r\right) +
\\&c e^{-i \theta}\left(  \frac{d}{dr}- \frac{\kappa}{r}\right)
Q_{n\kappa}\left(  r\right)
  =\varepsilon_{n\kappa}P_{n\kappa}\left(  r\right) \, ,\\
&-c e^{-i \theta} \left(  \frac{d}{dr}+\frac{\kappa}{r}\right)
P_{n\kappa}\left(  r\right) +\\
&\left(  V\left(  e^{i \theta}  r\right)
-c^{2}\right)  Q_{n\kappa}\left(  r\right)
=\varepsilon_{n\kappa}Q_{n\kappa}\left(  r\right) \, .
\end{align*}
Here $P$ and $Q$ are the conventional large and small radial components of the
Dirac bi-spinors. For a point-like nucleus $V\left( e^{i \theta}  r\right) = -
e^{-i \theta} Z/r$. For the DHF potential, the dependence
is
more complicated as it involves $\theta$-dependent core orbitals and requires
a self-consistent solution.

We have implemented the CRM method for the DHF equation using the finite-basis
set technique. Technically, we employed an expansion over B-splines and the
dual-kinetic-balance method~\cite{BelDer08} to avoid the so-called
``spurious'' states.  Representative numerical results for the $s_{1/2}$ symmetry are shown
in Table~\ref{Tab:DHF-CRM}. The generated finite basis is suitable for feeding
into the CI+MBPT code. Analytically, the scaling (\ref{Eq:rthetaScaling}) does
not affect energies of the bound states, but the eigenvalues of the continuum
are rotated in the complex plane by $-2\theta$. Our numerical data somewhat
deviate from this trend; this is related to  incompleteness of the spectrum
for finite basis sets, a result known in the literature.

\begin{table}[htb]
\caption{Dirac-Hartree-Fock energies for C$^{3+}$ ion in the complex-rotation
method. Rotation angle $\theta=10^\circ$. The finite basis set (dual-kinetic balance B-spline basis set)
for the $s_{1/2}$
symmetry consists of 40 orbitals. Cavity radius is $R=45$ bohr. The
$\varepsilon_{n\kappa}< -m_e c^2$ part of the spectrum is not shown.}
\label{Tab:DHF-CRM}
\begin{center}
\begin{tabular}{lrr}
\hline\hline
 $n$ & $\Re(\varepsilon_{n\kappa})$ & $\Im(\varepsilon_{n\kappa})$ \\
\hline
1 &$ -14.42319872  $&   $   -7.09\times 10^{-6}$  \\
2 &$ -2.365899978  $&   $    6.48\times 10^{-8}$  \\
3 &$-0.9891728632  $&   $    2.61\times 10^{-6}$   \\
4 &$-0.5410495311  $&   $    8.00\times 10^{-7}$   \\
5 &$-0.3407231660  $&   $   -2.21\times 10^{-4}$   \\
6 &$-0.2339320652  $&   $   -9.78\times 10^{-4}$   \\
.. & ..            & ..    \\
21&$   519.7385848$&$       -197.5  $  \\
22&$   1033.037284$&$       -380.7  $  \\
23&$   2013.838233$&$       -715.3  $  \\
24&$   3833.246780$&$       -1296.1 $ \\
..  & ..           & .. \\
40 &$ 13970526.56 $&$  -2466750.25 $  \\
\hline\hline
\end{tabular}
\end{center}
\end{table}

So far we discussed the one-body problem. It is the many-body part of the
problem, where the CRM method becomes invaluable. When the scaling of the
many-body Hamiltonian $H^\mathrm{eff}$ is carried out, new, complex, discrete eigenvalues
of $H^\mathrm{eff}(\theta)$ appear in the lower half of the complex energy plane. These are
the complex values that one associates with the resonances:
\begin{equation}
   E_r = E_\mathrm{res} - i \Gamma^a/2 \, .
\end{equation}
Here $E_\mathrm{res}$ and $\Gamma^a$  are the position and the autoionizing
width of the resonance we are after. For a complete set, these complex
eigenvalues remain unaffected as $\theta$ is varied, while the continuum
moves. For a finite basis, the $E_r(\theta)$ trajectory in the complex plane
``pauses'' or ``kinks'' at the physical position of resonances.

To reiterate, by diagonalizing the complex-symmetric CRM-scaled Hamiltonian, we
find the $\theta$ trajectories of the eigenvalues in the complex energy plane
and deduce the positions and widths of the DR resonances. Notice that no efforts are needed for computing the true scattering states. To efficiently diagonalize the complex symmetric matrices characteristic of the CRM method, we adopted a specialized Davidson-type eigensolver ``JDQZ''~\cite{FokSleVor96}.

\section{Numerical example: Beryllium-like carbon}
\label{Sec:C3plus}
As an illustration of our CI+MBPT+CRM toolbox, we consider
DR of the Li-like carbon. The DR pathway is
\begin{eqnarray*}
e^{-}+C^{3+}\left( 1s^{2}2s\right) \rightarrow \left[ C^{2+}\left(
1s^{2}2p_{j_{b}}nlj\right) \right] ^{\ast \ast }\rightarrow \\
C^{2+}\left(
...\right) +\rm{photon} \, .
\end{eqnarray*}
Low-energy DR for C$^{3+}$ was the focus of a combined theory-experiment paper
\cite{ManDeWEng98}; we compare our results with the results of that work
below.

The calculations were carried out using the relativistic basis
set with 40 orbitals per partial wave.
Representative numerical results for the $s_{1/2}$ symmetry are shown
in Table~\ref{Tab:DHF-CRM}.
The correlation
operator $\Sigma$ was computed with partial waves up to  $l_\mathrm{max}=5$.
The CI model space was spanned by the 25 lowest-energy
virtual orbitals for each partial wave up to $l_\mathrm{max}=4$.
For example, for the $J=1$, odd symmetry we include all possible
11 angular channels. The most important channel is represented by the
 $np_{j} n's_{1/2}$ and $nd_{j} n'p_{j'}$
configurations. The $nd_{j} n'p_{j'}$ configurations give rise to leading
contributions to the low-energy $2p4d$ DR resonances. These resonances are
embedded into the $ 2s_{1/2} \epsilon p_{j}$ continuum. Including both types of
configurations allows us to incorporate the interaction of the DR resonances with
the continuum to all orders of MBPT and avoid accidental degeneracies.  While the theoretical formulation of  Ref.~\cite{ManDeWEng98} is similar to
ours, their model space excludes the $2snp$ continuum and its effect is
taken into account in all-order MBPT.

\begin{figure}[tbh]
\begin{center}
\includegraphics*[scale=0.75]{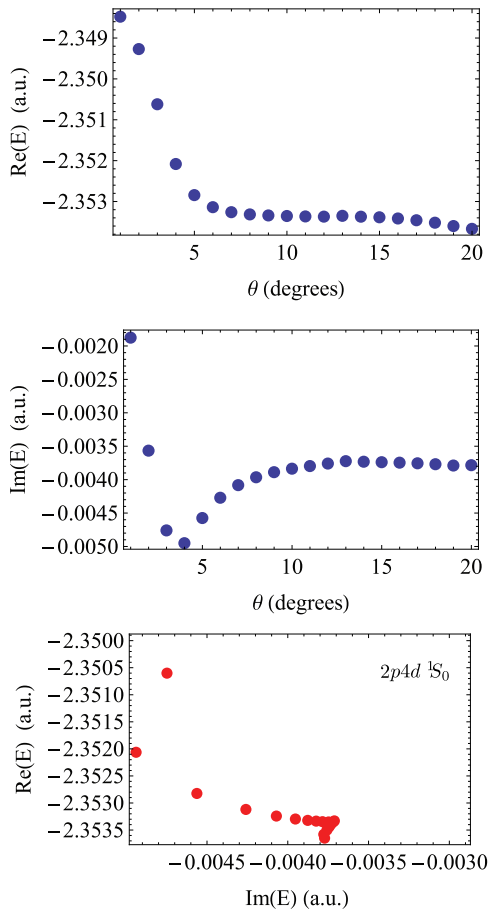}
\end{center}
\caption{(Color online) Illustration of the CI+MBPT+CRM method for locating the  $2p4d\, ^1\!S_0$ resonance. The upper (central) panel shows dependence
of the real (imaginary) part of the energy on the CRM rotation angle $\theta$. The bottom panel
shows the corresponding trajectory in the complex plane. The kink in the trajectory marks the position
of the resonance.
 \label{Fig:CRMtrajectory}}
\end{figure}

Calculations of the CRM trajectories of energy levels were carried out for angles $0^\circ-30^\circ$ with a step of $1^\circ$. An example of a trajectory for the $2p4d\, ^1\!S_0$  odd-parity resonance
is shown in Fig.~\ref{Fig:CRMtrajectory}.

Our numerical results are compiled in Table~\ref{Tab:Sets}. We tabulate energies, widths, and strengths of DR resonances falling within a
0.6 eV range above the $1s^2 2s$ threshold.
In this table, we also compare our values with the previous theoretical results of
Ref.~\cite{ManDeWEng98}. The energies are also compared with the NIST recommended values~\cite{NIST_ASD}. We find that the overall agreement for energy positions is excellent and does not deviate by more
than 2-3 meV. Detailed consideration reveals that the NIST recommended values for the position of the
$2p4d\,^3\!P_J$ resonance differ both from our and Ref.~\cite{ManDeWEng98} predictions by as much as 13 meV.  Even more strikingly, both our and Ref.~\cite{ManDeWEng98} predictions disagree with the NIST recommended value
by a very large value of 147 meV  for the position of the $2pd4\, ^1\!F_3$ resonance.

Among 21 tabulated resonance positions, there is only one large disagreement between our work and theory of Ref.~\cite{ManDeWEng98}: this happens for the $2p4d\, ^1\!S_0$ resonance where the two calculations differ by  34 meV. Considering an excellent agreement for other 20 resonances between the two calculations, such a disagreement may indicate a typographical mistake in Ref.~\cite{ManDeWEng98}.
Ref.~\cite{ManDeWEng98} reports the following experimental positions of the resonances: 0.182 eV for $2p4d\, ^3D$, 0.244 eV for
the unresolved $2p4f\, ^{1,3}\!F$, 0.438 eV for $2p4f\, ^3\!D$, and
0.578 eV for $2p4d\, ^1\!P$ with experimental uncertainty of 5 meV. All of these values are in agreement
with our theoretical predictions.

A comparison of resulting
autoionizing widths between our and Ref.~\cite{ManDeWEng98} calculations indicates a good agreement
for broad resonances. The agreement for narrow ( width $< 0.5$ meV) is less satisfying. Experimentally,  the width of such
resonances is determined by the experimental convolution function and no
definitive conclusions can be drawn on the basis of theory-experiment
comparison. Notice, that
the $2p4d ^3D$ and $2p4f ^{1,3}F$ resonances were resolved in the experiment~\cite{ManDeWEng98}. The calculated rate in Ref.~\cite{ManDeWEng98} was larger than the experimental one by 50\%. It was
not clear whether  the source of the disagreement was theory or experiment.
Perhaps, our disagreement for the width of narrow resonances with theory~\cite{ManDeWEng98} may indicate enhanced sensitivity to details of theoretical treatment.
While there are discrepancies for the autoionizing widths of the narrow resonances,
our CI+MBPT+CRM strengths of the resonances compare well with calculations of  Ref.~\cite{ManDeWEng98} (see Table~\ref{Tab:Sets}).

\section{Conclusion}
To summarize, we report developing a new method for computing properties of
low-energy resonances in dielectronic recombination. A high-precision description of low-energy resonances
is particularly challenging as it is sensitive to correlations. At the same time, uncertainties
in the positions of the resonances drastically affect practically important recombination rates in low-temperature plasmas.
Our theoretical approach is based on
combining configuration-interaction method with many-body perturbation theory and complex rotation method (CI+MBPT+CRM). The method is {\em ab initio} relativistic. To gauge the
accuracy of the developed CI+MBPT+CRM approach, we computed low-energy resonances in Be-like carbon.
We find a good agreement with the earlier high-precision study by \citet{ManDeWEng98}.
While here we studied a divalent ion,
our developed methodology and computational toolbox is well suited for exploring resonances
in systems with several valence electrons outside a closed-shell core.

\emph{Acknowledgements ---}
  We would like to thank D. Savin and E. Lindroth for discussions and
  Frank Greenhalgh for comments on the manuscript.
  This work was supported in part by the US NSF and by the US NASA under Grant/Cooperative
  Agreement No. NNX07AT65A issued by the Nevada NASA EPSCoR program.

\begin{table*}[htb]
\caption{ {\em Ab initio} values of energies (relative to the $1s^22s \
  ^2$S$_{1/2}$ threshold), the widths and strengths of resonances  within 0.6 eV
above the ionization limit of C~III. The computed values are  compared with theoretical
results of Mannervik {\em et al.,} (see Table~I in Ref.~\cite{ManDeWEng98}) and NIST 
recommended energies \cite{NIST_ASD}.}
\label{Tab:Sets}
\begin{ruledtabular}
  \begin{tabular}{l l c l c c c c c c}
&&&\multicolumn{3}{c}{Energy (eV)} &
\multicolumn{2}{c}{Width (meV)} &
\multicolumn{2}{c}{Strength (10$^{-20}$ eV cm$^2$)} \\

\multicolumn{2}{c}{Term} &
\multicolumn{1}{c}{$J$} &
\multicolumn{1}{c}{Present} &
\multicolumn{1}{c}{NIST\cite{NIST_ASD}} &
\multicolumn{1}{c}{Ref.\cite{ManDeWEng98}} &
\multicolumn{1}{c}{Present} &
\multicolumn{1}{c}{Ref.\cite{ManDeWEng98}} &
\multicolumn{1}{c}{Present} &
\multicolumn{1}{c}{Ref.\cite{ManDeWEng98}} \\

\hline

%
%
%
%
%
%

 $2p4d$ & $^3$D$^o$ & 1 & 0.179 &0.181 & 0.176 & 0.1  & 0.09 &  10.3 & 11.2 \\    
        &           & 2 & 0.183 &0.181 & 0.177 & 0.005& 0.18 &  17.2 & 18.5 \\    
        &           & 3 & 0.183 &0.181 & 0.180 & 0.005& 0.08 &  22.5 & 25.4 \\   

 $2p4f$ & $^1$F     & 3 & 0.238 &      & 0.236 & 0.45 & 0.10 &   5.7 &  6.0 \\  
        & $^3$F     & 2 & 0.242 &0.239 & 0.240 & 0.05 & 0.001&   3.8 &  3.9 \\ 
        &           & 3 & 0.243 &0.241 & 0.242 & 0.6  & 0.25 &   5.3 &  5.5 \\   
        &           & 4 & 0.245 &0.245 & 0.243 & 0.4  & 0.33 &   6.8 &  7.1 \\   

 $2p4d$ & $^3$P$^o$ & 0 & 0.291 &0.279 & 0.292 &  50  &  52  &   1.1 &  1.3 \\   
        &           & 1 & 0.289 &0.279 & 0.289 &  50  &  52  &   3.2 &  3.7 \\   
        &           & 2 & 0.284 &0.279 & 0.285 &  50  &  52  &   4.7 &  5.8 \\   
 $2p4f$ & $^3$G     & 3 & 0.353 &      & 0.351 & 117  & 115  &   3.1 &  3.3 \\   
        &           & 4 & 0.356 &      & 0.353 & 117  & 115  &   3.9 &  4.2 \\   
        &           & 5 & 0.365 &      & 0.360 & 117  & 115  &   4.7 &  5.1 \\   
        & $^1$G     & 4 & 0.379 &      & 0.375 & 118  & 115  &   3.1 &  3.2 \\   
        & $^3$D     & 1 & 0.435 &0.433 & 0.433 & 1.1  & 1.01 &   1.2 &  1.2 \\   
        &           & 2 & 0.431 &0.432 & 0.430 & 1.1  & 1.01 &   2.0 &  2.0 \\   
        &           & 3 & 0.427 &0.425 & 0.426 & 1.2  & 1.01 &   2.8 &  2.8 \\   
        & $^1$D     & 2 & 0.451 &      & 0.452 & 0.6  & 0.22 &   1.8 &  1.8 \\   
 $2p4d$ & $^1$F$^o$ & 3  & 0.461& 0.314 & 0.460& 232   & 236 &   7.4 &  8.5 \\  
 $2p4p$ & $^1$S     & 0  & 0.451&       & 0.485& 202   & 221 &   0.2 &  0.2 \\  
 $2p4d$ & $^1$P$^o$ & 1  & 0.583&       & 0.586&  44   &  46 &   1.9 &  2.0 \\   
\end{tabular}
\end{ruledtabular}
\end{table*}


\end{document}